\documentclass[a4paper,twocolumn]{revtex4}

\usepackage{amsmath}
\usepackage{graphicx}
\usepackage[english]{babel} 
\usepackage[latin1]{inputenc} 
\usepackage[T1]{fontenc}

\begin{document}

\title{Entropy of chains placed on the square lattice}
\author{Wellington G. \surname{Dantas}}
\email{wgd@if.uff.br}  
\author{J\"urgen F. \surname{Stilck}}
\email{jstilck@if.uff.br}  
\affiliation{Instituto de F\'{\i}sica, Universidade Federal Fluminense, \\
Campus da Praia Vermelha,  \\
Niter\'oi, RJ, 24.210-340, Brazil.}

\date{\today}

\begin{abstract}
We obtain the entropy of flexible linear chains composed of $M$ monomers
placed on the square lattice using a transfer matrix approach. An excluded
volume interaction is included by considering the chains to be self-and
mutually avoiding, and a fraction $\rho$ of the sites are occupied by
monomers. We solve the problem exactly on stripes of increasing width 
$m$ and then extrapolate our results to the two-dimensional limit $m \to
\infty$ using finite-size scaling. The extrapolated results for several
finite values of $M$ and in the polymer limit $M \to \infty$ for the cases
where all lattice sites are occupied ($\rho=1$)  and for the partially filled
case $\rho<1$ are compared with earlier results. These results are exact for
dimers ($M=2$) and full occupation ($\rho=1$) and derived from series
expansions, mean-field like approximations, and transfer matrix calculations
for some other cases. For small values of $M$, as well as for the polymer
limit $M \to \infty$, rather precise estimates of the entropy are obtained.
\end{abstract}

\maketitle

\section{Introduction}

The term dimer was introduced in the thirties \cite{fr37} as an abbreviation
for diatomic molecules in a model for their adsorption on crystal
surfaces. Later dimer models were applied in the study of other physical
systems such as ferroelectrics, and much is known about their thermodynamic
properties \cite{nyb89}. One relevant question in these models is the entropy
associated to placing dimers on a regular lattice. For the particular case of
full covering of the square lattice by dimers, this question was answered
exactly some time ago, using a technique based on pfaffians
\cite{fis,kast,fisetemp}. However, even the generalization of this problem for
the case on partial covering of the square lattice is still an open question
today, no exact result being known. 

In this paper we address a generalization of the entropy of dimers problem,
considering entropy related to covering the square lattice with chains with
$M$ monomers each (we will call them $M$-mers), as a function of the fraction
$\rho$ of sites of the lattice occupied by monomers. The chains will be
considered flexible, so that there is no energy associated to bending
them. Since the only energy in the model is the infinite excluded volume
interaction, which forbids the presence of more than one monomer on the same
lattice site, the problem is athermal. It may be a simple model for the
adsorption of monodisperse flexible chains on the surface of a
crystal. Besides the exact solution of the problem for $M=2$ and $\rho=1$
mentioned above, other cases were already considered in the literature. Rather
precise transfer matrix calculations were performed in the polymer limit $M
\to \infty$ for hamiltonian walks ($\rho=1$) \cite{dup}. There are also
mean-field approximations \cite{flor}, Bethe - and Husimi lattice calculations
\cite{jsmo}, and series expansions in $q^{-1}$, where $q$ is the number of  
first neighbors of each site in the lattice \cite{nem}, and in those 
calculations approximate values for the entropy are obtained for both the 
full ($\rho=1$) and partial ($\rho<1$) coverage cases.

In this paper we obtain estimates for the entropy of flexible
$M$-mers placed on the square lattice, using transfer matrix techniques. This
is done calculating numerically exact values for the entropy of the problem on
strips with finite widths $m$ and periodic boundary conditions and then using
finite-size scaling to extrapolate the results to the two-dimensional limit 
$m \to\infty$. We separate the problem in cases where $M$ is finite or infinite
(polymer limit). Also, the case of full coverage ($\rho=1$) may be treated
separately from the general case. In the general case, it is convenient to
address the problem in an ensemble which is grand-canonical with respect to
the number of monomers placed on the lattice, whereas for full coverage it is
easier to perform a microcanonical calculation.

The expressions we used to calculate the entropy are shown in section
\ref{entro}. The model is discussed in more detail and the transfer matrices
are described in section \ref{tm}. Our results for the entropies may be found
in section \ref{nr}, as well as the extrapolation procedure and their
results. Section \ref{conc} presents final discussions and conclusions.

\section{Determination of the entropy}
\label{entro}

For the case of full coverage, it is convenient to obtain the entropy directly
from Boltzmann's expression
\begin{eqnarray}
\label{erc}
s(\rho=1) = \lim_{N\rightarrow\infty}\frac{S}{Nk_B} = 
\lim_{N\rightarrow\infty}\frac{1}{N}\ln{\Omega},
\end{eqnarray}
where $\Omega$ is the number of ways to fill the lattice with $N$ sites
completely with $M$-mers. In the polymer limit $M \to \infty$, we consider a
{\em single} hamiltonian walk, that is, a self-avoiding walk (SAW) which visits all
sites of the lattice.

In the general case where a fraction $\rho$ of lattice sites is occupied by
monomers, we define the grand-canonical partition function
\begin{eqnarray}
\label{gfp}
\Xi(z) = \sum_{p}z^{pM}\Gamma(M,N,p),
\end{eqnarray}
where $z$ is the activity of a monomer and $\Gamma(M,N,p)$ is the number of
ways to place $p$ chains with $M$ monomers each on the lattice with $N$
sites. For the polymer limit, again a single SAW is placed on the lattice and
the partition function is defined as 
\begin{eqnarray}
\label{gfpp}
\Xi(z) = \sum_{n}z^{n}\Gamma(n,N),
\end{eqnarray}
where $\Gamma(n,N)$ is the number of ways to place a SAW with $n$ monomers on
the $N$-site lattice. The density of monomers may now be written as
\begin{eqnarray}
\label{den}
\rho(z) = z\frac{d}{dz}\phi(z),
\end{eqnarray}
where the thermodynamic potential per lattice site is defined as
\begin{eqnarray}
\label{phi}
\phi(z) = \lim_{N\rightarrow\infty}\frac{1}{N}\ln{\Xi(z)}.
\end{eqnarray}
In the thermodynamic limit, a Legendre transformation allows us to rewrite the
potential as
\begin{eqnarray}
\label{rel}
\phi(z)\sim \max_{\rho}\{\rho \ln{z}+s(\rho)\},
\end{eqnarray}
and thus the entropy may be written as
\begin{eqnarray}
\label{entden}
s(\rho) = -\int_{0}^{\rho} \ln{z}(\rho')d\rho',
\end{eqnarray}
with $s(0) = 0$.

\section{Definition of the transfer matrix}
\label{tm}

\begin{figure}[h!]
\begin{center}
\includegraphics[height=3.0cm]{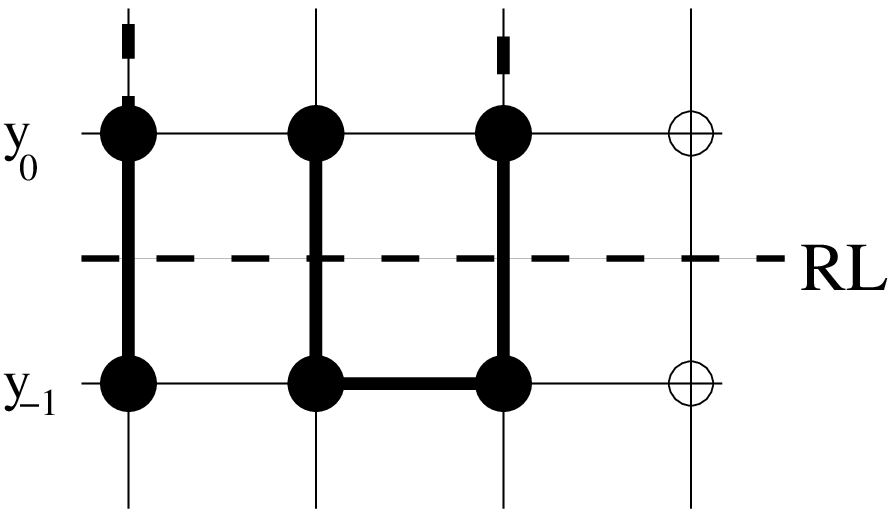}
\caption{Example of a state for $m=4$ vertical bonds. The reference line 
indicates the set of vertical bonds whose configuration is described.}
\label{ex1}
\end{center}
\end{figure}

\begin{figure}[h!]
\begin{center}
\includegraphics[height=3.1cm]{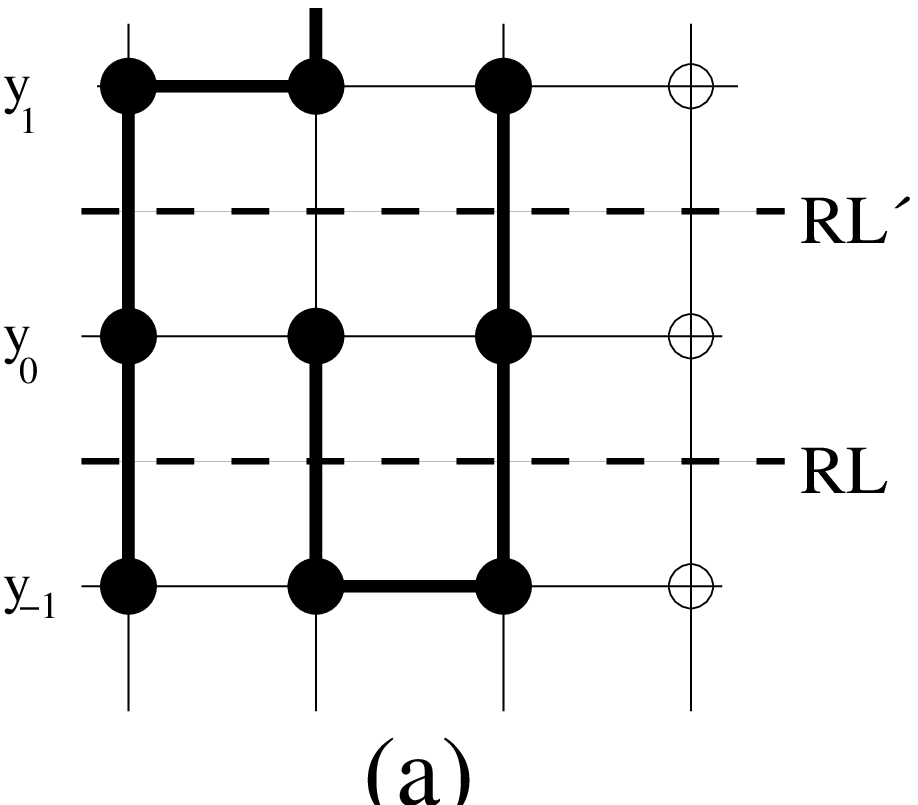}
\includegraphics[height=3.1cm]{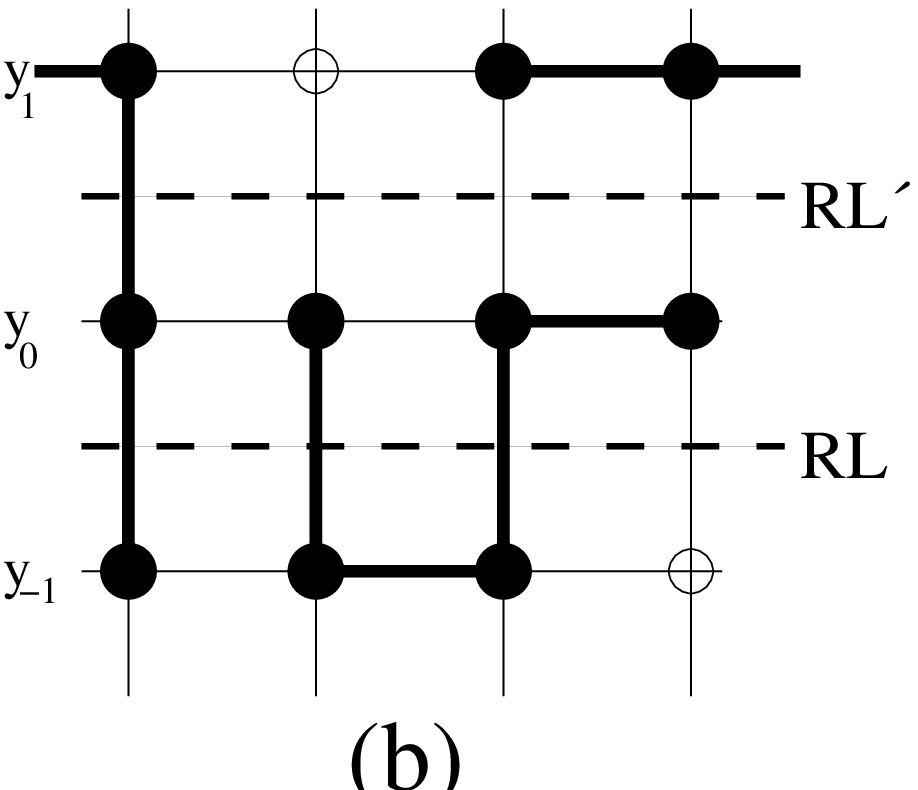}
\caption{Possible continuations (reference line RL') following the
configuration depicted in \ref{ex1} (reference line RL).}
\label{ex2}
\end{center}
\end{figure}

We proceed defining a strip of width $m$ on the square lattice in the $(x,y)$
plane, so that $1 \leq x \leq m$ and $-\infty \leq y \leq \infty$, with
periodic boundary conditions in both directions. A transfer matrix may be
built for this problem, inspired on the prescription due to Derrida \cite{der}
for infinite chains in strips. We thus consider the operation of including an
additional 
step to the strip in the positive $y$ direction,adding $m$ new sites of the
lattice. To properly take into account the statistical weight of the new step,
we may define the state of the $m$ vertical bonds of the lattice which are
incident to the new sites by specifying:

\begin{enumerate}

\item The number of monomers already present in the chain which passes through
the vertical bond (it is equal to 0 if no chain is present). These numbers may
be put into a vector $|p\rangle$, with $m$ components. It is necessary to keep
track of this information so that we know when to end each chain.

\item The pairs of bonds which are connected to each other through a path
lying entirely below the reference line (see figure \ref{ex1}). 
These pairs may be also be specified by a $m$-component vector
$|v\rangle$, associating a different positive integer to each pair of
connected bonds and 0 to the ones which are not connected to any other. This
connectivity information prevents us from closing a ring at any level, since
this configuration is not allowed in the model.

\end{enumerate}
As an example of these definitions, we consider the case of pentamers ($M=5$)
placed on a strip of width $m=4$. Among the possible configurations of a set
of vertical bonds, the one depicted in figure \ref{ex1} is described by the vectors
$|v\rangle = (0,1,1,0)$ and $|p\rangle = (1,2,2,0)$. Elements of the line
associated to this state of the transfer matrix $\mathcal{T}$ are obtained
considering the possible continuations of the state $\{|v\rangle,|p\rangle\}$
one step upwards, as shown in two examples in figure \ref{ex2}. The resulting
state in figure \ref{ex2}.a is described by the vectors $|v\rangle =
(0,0,0,0)$ and $|p\rangle = (2,0,4,0)$, while the final state in figure
\ref{ex2}.b corresponds to the vectors $|v\rangle = (0,0,0,0)$ and
$|p\rangle = (2,0,0,0)$. 
Each monomer placed on a site between the two sets of
vertical bonds contributes with an activity $z$ to the partition function, so
that the element of the transfer matrix which corresponds to the first
configuration is equal to $z^3$, while the second configuration is associated
to an element equal to $z^4$ in the transfer matrix. Only the second
configuration contributes in the case of full occupancy. 

For the polymer case $M \to \infty$, a single chain passes through the whole
strip, so that it is enough to describe the connectivity at a particular set
of $m$ vertical bonds by indicating the bond which is connected to the initial
monomer of the chain (in $y \to -\infty$) and the pairs of bonds connected to
each other, exactly as was done in the original work of Derrida
\cite{der}. Thus, a single vector $|v\rangle$ is enough to describe the
state in this limit. 

Once the transfer matrix $\mathcal{T}$ is obtained, the entropy of the model
on the strip in the thermodynamic limit is related to the largest eigenvalue
of the matrix. For the case of full occupancy ($\rho=1$), the number of
configurations is given by
\begin{eqnarray}
\Omega = Tr(\mathcal{T'}^{l}),
\end{eqnarray}
where $N=ml$ is the number of sites and the elements of the matrix
$\mathcal{T'}$ are defined by the limit
\begin{eqnarray}
\mathcal{T'}_{i,j} = \lim_{z\rightarrow\infty}\frac{\mathcal{T}_{i,j}}{z^m}.
\end{eqnarray}
The entropy is then related to the largest eigenvalue $\lambda^{'}$ of this
matrix, so that 
\begin{eqnarray}
s(\rho=1)=\frac{1}{m}\ln\lambda^{'}.
\end{eqnarray} 

For the general case, where a fraction $\rho$ of lattice sites are occupied by
monomers, the grand-canonical partition function is related to the transfer
matrix through  
\begin{eqnarray}
\Xi(z) = Tr(\mathcal{T}^l),
\end{eqnarray}
and thus the density $\rho(z)$ will be
\begin{eqnarray}
\rho(z) =  
\lim_{N\rightarrow\infty}\frac{z}{N}\frac{d}{dz}ln\Xi(z) =
\frac{z}{m}\frac{d}{dz}ln\lambda,
\label{rho}
\end{eqnarray}
where $\lambda$ is the largest eigenvalue of the transfer matrix
$\mathcal{T}$. 
This relation may be inverted to obtain the entropy as a function of the
density using equation \ref{entden}.

\section{Numerical results}
\label{nr}

\begin{figure}[h!]
\begin{center}
\includegraphics[height=5.0cm]{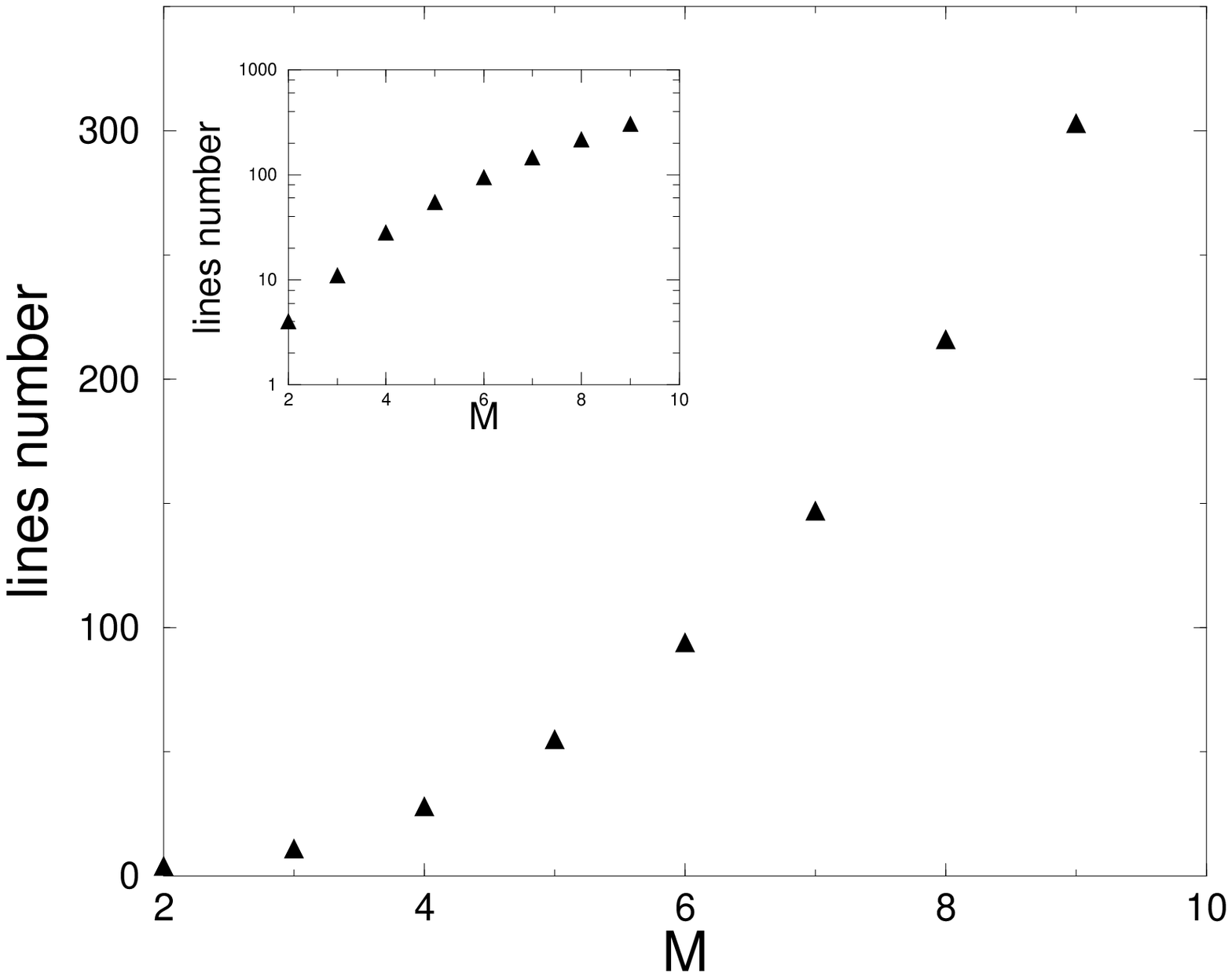}
\includegraphics[height=5.0cm]{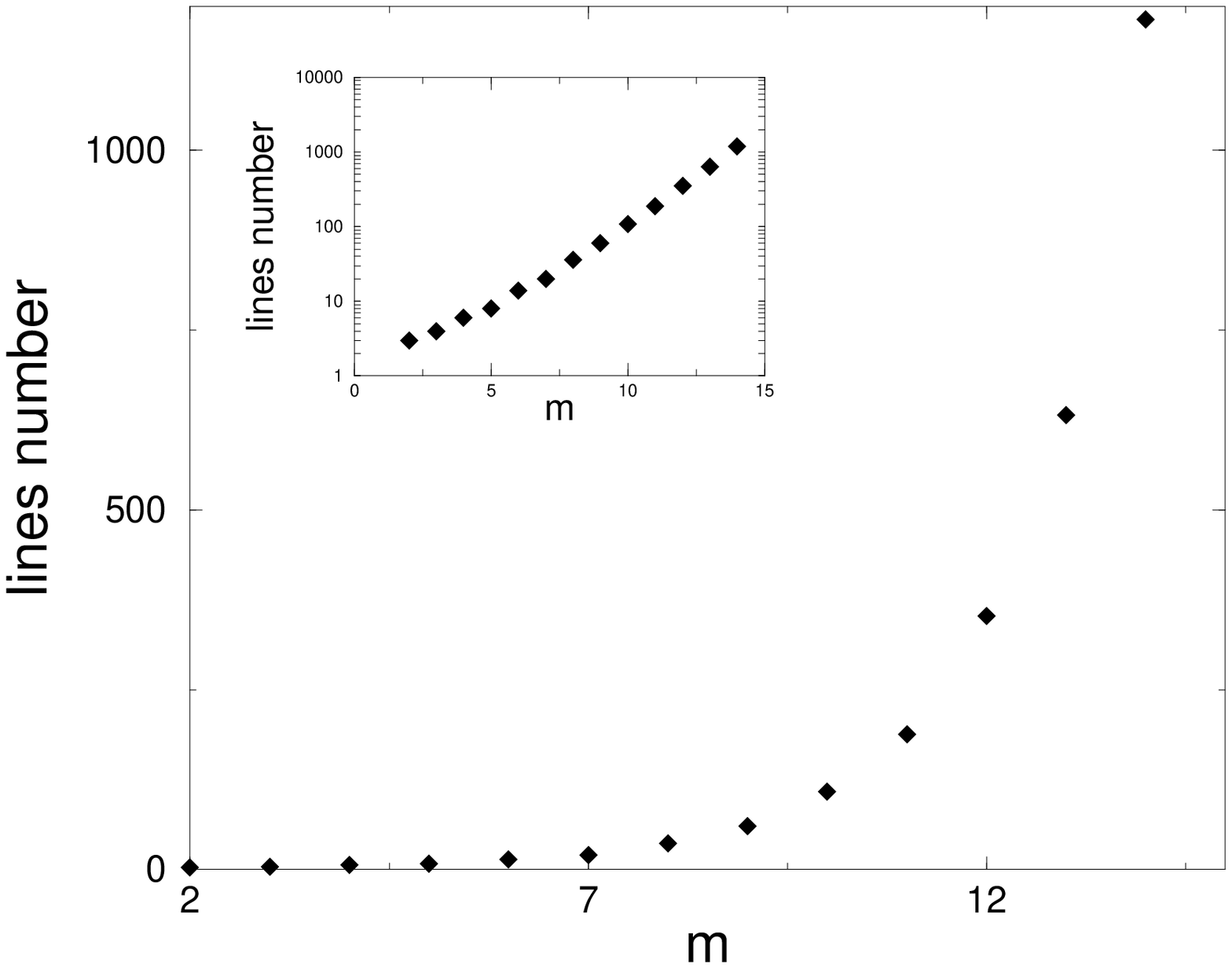}
\caption{Dimension of the transfer matrix as a function of the molecular
weight $(m=2)$ and width of the strip $(M=2)$, respectively.}
\label{cres}
\end{center}
\end{figure}

The size of the transfer matrix increases very fast with both the molecular 
weight $M$ and the width $m$ of the strip grow. This sets an upper limit to 
the widths we were able to consider for each chain of a given molecular 
weight. Figure \ref{cres} shows this effect, and in the inset one may 
appreciate that the growth of the transfer matrix is roughly exponential.

Furthermore, as was already observed in similar calculations for polymers
\cite{hite}, the values of the entropy for each class of $M$-mers are split
into subsets with different finite-size scaling behavior in each subset
according to the width of the strips, so that extrapolations must 
be done within each subset. These splitting seems to be related to frustration
effects in the limit of the fully occupied lattice and the subsets
are indicated in table \ref{sep}, where all the widths we considered are
given. 

\begin{table}[h!] 
\caption{Entropies calculated for each $M$-mer, divided in subsets with the same
finite-size scaling behavior
\label{sep}}
\begin{center}
\begin{tabular}{c c}  \hline\hline
Molecular weight & Entropies of each subset \\ \hline\hline
$M=2$  & \\ 
       &    $\{s'_1\} = \{s_2,s_4,s_6,...,s_{14}\}$ \\
       &    $\{s'_2\} = \{s_3,s_5,s_7,...,s_{13}\}$ \\ \hline 

$M=3$  &   \\ 
       &    $\{s'_1\} = \{s_3,s_6,s_9,s_{12}\}$ \\
       &    $\{s'_2\} = \{s_2,s_4,s_5,s_7,s_8,s_{10},s_{11}\}$\\ \hline 
$M=4$  & \\  
       &   $\{s'_1\} = \{s_4,s_8\}$ \\ 
       &   $\{s'_2\} = \{s_2,s_6,s_{10}\}$ \\
       &   $\{s'_3\} = \{s_3,s_5,s_7,s_9\}$ \\ \hline
$M=5$  & \\  
       &   $\{s'_1\} = \{s_5\}$ \\ 
       &   $\{s'_2\} = \{s_2,s_3,s_{4},s_6,s_7,s_8\}$ \\ \hline
$M=6$  & \\         
       &   $\{s'_1\} = \{s_6\}$ \\ 
       &   $\{s'_2\} = \{s_3\}$ \\
       &   $\{s'_3\} = \{s_2,s_4\}$ \\ 
       &   $\{s'_4\} = \{s_5,s_7\}$ \\ \hline
$M=7$  & \\  
       &   $\{s'_1\} = \{s_2,s_3,s_4,s_5,s_6\}$ \\ \hline
$M=8$  & \\  
       &   $\{s'_1\} = \{s_4\}$ \\ 
       &   $\{s'_2\} = \{s_2\}$ \\
       &   $\{s'_3\} = \{s_3,s_5\}$ \\ \hline 
$M=9$  & \\         
       &   $\{s'_1\} = \{s_3\}$ \\ 
       &   $\{s'_2\} = \{s_2,s_4\}$ \\ \hline  
$M\rightarrow\infty$  & \\ 
       &    $\{s'_1\} = \{s_2,s_4,s_6,...,s_{12}\}$ \\
       &    $\{s'_2\} = \{s_3,s_5,s_7,...,s_{13}\}$ \\ 
\hline\hline 
\end{tabular}
\end{center}
\end{table}

The data of each subset were extrapolated to the two-dimensional limit $m \to
\infty$ using the Shanks transformation \cite{barb}, since we expect
finite-size corrections to be exponential. Since at least three values for the
entropy are needed in a subset to obtain an estimate for the two-dimensional
entropy and its confidence interval, not all subsets may be extrapolated, so
that, for example, no estimate could be found for hexamers ($M=6$), where we
calculated entropies for widths up to $m=7$. The final estimate was chosen to
be the highest possible extrapolant, and the error associated to it was
obtained from the previous generation of extrapolants, through
\begin{eqnarray}
\epsilon = \lim_{l'\rightarrow\infty} 2|s_{l'-1}-s_{l'+1}|.
\end{eqnarray}
The extrapolated values of the entropies for $\rho=1$ and their uncertainties
are displayed in table \ref{extrap}, together with values obtained with other
techniques and best values found in the literature.

\begin{table*}
\caption{Entropy of $M$-mers on the square lattice for full coverage
$(\rho=1)$ obtained through different techniques
\label{extrap}}
\begin{center}
\begin{tabular}{c c c c c c}  \hline\hline
$M$& MF & Bethe & Series & Transfer Matrix & Best value\\ 
\hline\hline
$2$ &             &            &             &         \\
                  & 0.19315    & 0.26162     & 0.26867 & 0.29120$\pm$ 0.00071&

0.29156
\\ \hline
$3$ &            &             &             &            \\         
                   &  0.39268    & 0.42284     & 0.41699  & 0.41201$\pm$ 0.00002&
\\ \hline       
$4$ &            &             &           &      \\
                   &  0.46301    & 0.48166     & 0.48889  & 0.51486$\pm$ 0.0045&
\\ \hline
$5$ &            &             &          &        \\
                   & 0.49229     & 0.50669     & 0.51008  & 0.49917$\pm$ 0.00091
\\ \hline
$7$ &            &             &   &               \\
                 & 0.51008     & 0.52217     & 0.52170  & 0.54770$\pm$ 0.15301&
\\ \hline
$M\rightarrow\infty$ & & &\\
&0.3863 & 0.4055 & 0.4090 &0.3870$\pm$ 0.0009&0.3866
\\ \hline\hline
\end{tabular}
\end{center}
\end{table*}

Our results may be compared with other values in the
literature. One may notice that the mean-field estimates are systematically
smaller than the values obtained here, but no such general trend is apparent 
for the Bethe- and Husimi lattice results. The estimate for dimers agrees 
with the exact value obtained by Fisher, Temperley and Kasteleyn
\cite{fis,kast,fisetemp}, and the entropy for hamiltonian walks $(M \to
\infty)$ is consistent with both transfer matrix calculations  \cite{dup} and
the result of series expansions up to third order in $q^{-1}$ \cite{lise},
which is $s_{\infty}\approx0.38629$.

Another relevant question is the value of the molecular weight which maximizes
the entropy at full occupancy. Mean-field and Bethe lattice results show
maximum entropy at $M=8$ for a lattice with coordination number $q=4$ 
\cite{flor,jsmo}, while series up to second order in 
$q^{-1}$ on the square lattice result in a maximum entropy at $M=7$. Our
results suggest that this maximum actually occurs at $M=4$ on the square
lattice, if we suppose that that only one maximum exists in the curve
$s(\rho=1) 
\times M$ and also disregard the value obtained for $M=7$ due to the large
uncertainty associated to it. 

\begin{figure}[h!]
\begin{center}
\includegraphics[height=6.5cm]{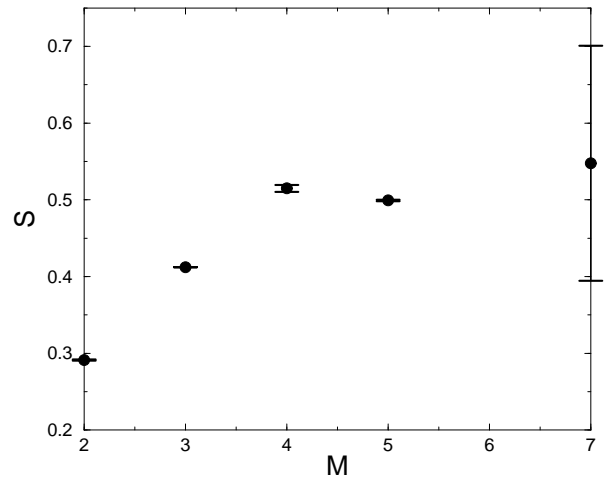}
\caption{Entropy at full occupancy of the lattice as a function of the
molecular weight $M$}
\label{maxm}
\end{center}
\end{figure}

\begin{figure}[h!]
\begin{center}
\includegraphics[height=6.5cm]{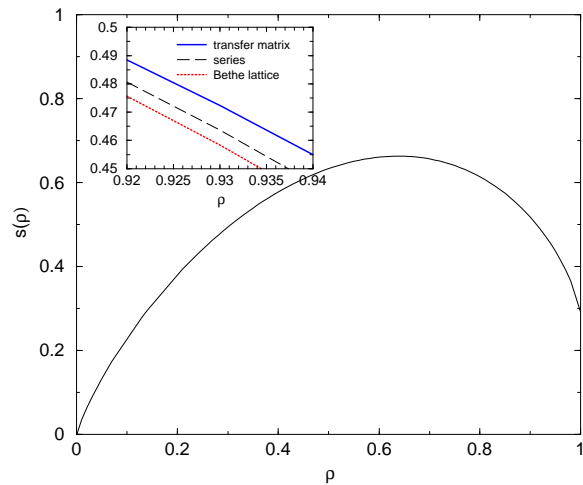}
\caption{Entropy for dimers as a function of the density. In the inset,
results of values obtained in this work are compared with results from series
expansion \cite{nem} and Bethe lattice \cite{jsmo} calculations.}
\label{rpp}
\end{center}
\end{figure}

For partial occupancy of the lattice, the results are similar to the ones
shown in figure \ref{rpp} for the entropy of dimers as a function of the
fraction of occupied lattice sites $\rho$. For all the cases we considered
the entropy displays a single maximum. The density at which this maximum occurs
increases with $M$, getting closer to $\rho \approx 0.79$, the value found in
the polymer limit. The densities and maximum entropies are listed in table
\ref{maxs}, and it may be noticed that the largest value for the maximum
entropy occurs for tetramers, as was also found for $\rho=1$. 

\begin{table}[h!] 
\caption{Maximum values of the entropy as a function of the density.}
\label{maxs}
\begin{center}
\begin{tabular}{c c c}  \hline\hline
$M$& density of maximum entropy & maximum entropy \\ \hline\hline
$2$ & 0.64 & 0.66 \\
$3$  & 0.71 & 0.70 \\
$4$  & 0.76 & 0.74\\
$5$  & 0.76 & 0.73\\
$7$  & 0.78 &0.72\\
$M\rightarrow\infty$ & 0.79 &0.56
\\ \hline\hline
\end{tabular}
\end{center}
\end{table}

\begin{figure}[h!]
\begin{center}
\includegraphics[height=6.5cm]{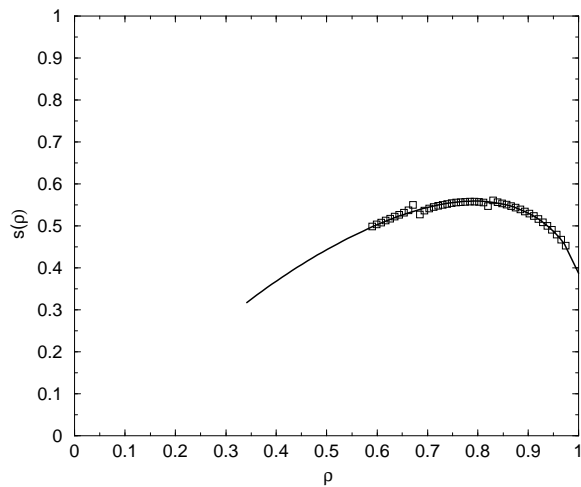}
\caption{Entropy of polymers as a function of the density. The squares
correspond to extrapolations of the values on strips of finite widths and the
full line is the result for a strip of width $m=7$.}
\label{polex}
\end{center}
\end{figure}

In the polymer limit $M \to \infty$, the model of a polymer placed on a strip
in the grand-canonical ensemble displays a first order phase transition at a
critical activity $z_c$, with the coexistence of a non-polymerized phase
($\rho=0$) and a polymerized phase ($\rho=\rho_c>0$) \cite{pw83}. As the width $m$ of
the strip is increased, the discontinuity in the density at the transition
becomes smaller and in the two-dimensional limit $m \to \infty$ a continuous
transition is found at $z_c \approx 0.3790522$ \cite{gut}. The entropy of a
polymer on a strip of finite width is therefore not defined for $\rho<\rho_c$,
as may be seen in figure \ref{polex}, where the extrapolated value of the
entropy as a function of the density for polymers is depicted. As larger
widths are considered, the step in the entropy decreases and in the
two-dimensional limit we have $s(\rho=0)=0$.

\section{Conclusion}
\label{conc}

In this paper we estimate the entropy of chains with $M$ monomers each placed
on the square lattice as a function of the fraction $\rho$ of lattice sites
occupied by monomers. The estimates were obtained by extrapolating numerically
exact values for the entropy on strips of finite widths $m$, calculated using
a transfer matrix approach, to the two-dimensional limit $m \to \infty$. 

\begin{figure}[h!]
\begin{center}
\includegraphics[height=6.5cm]{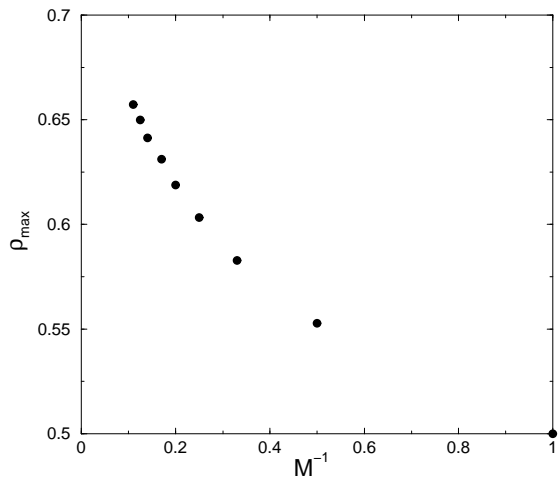}
\vspace*{-0.2cm}
\caption{Density of maximum entropy as a function of $1/M$.}
\label{maximum}
\end{center}
\end{figure}

A particular case where results may easily be obtained analytically is the one
dimensional problem ($m=1$). In this case, the dimension of the transfer
matrix $\mathcal{T}$ is equal to $M$, and we have
\begin{equation}
\mathcal{T}_{i,j}=\delta_{i,1}\delta_{j,1}+
z(\delta_{i+1,j}+\delta_{i,M}\delta_{j,1}),
\end{equation}
where $1 \leq i,j \leq M$. One may easily obtain the secular equation of this
matrix, which is
\begin{equation}
\lambda^M-\lambda^{M-1}-z^M=0.
\end{equation}
The density may then be found as a function of the largest eigenvalue
$\lambda$ through using the secular equation above and also expression
\ref{rho}. One obtains
\begin{equation}
\lambda=\frac{1-\alpha\rho}{1-\rho},
\end{equation}
where $\alpha=(M-1)/M$. Then the entropy may be found by performing the
integration in equation \ref{entden} changing the integration variable from
$\rho$ to $\lambda$. The result is
\begin{eqnarray}
s&=&(1-\alpha\rho)\ln(1-\alpha\rho)-(1-\rho)\ln(1-\rho)+\nonumber\\
&-&\rho(1-\alpha)\ln[\rho(1-\alpha)].
\end{eqnarray}
This result is equal to the expression which is obtained if the coordination
number of the Bethe lattice result (expression 22 in \cite{jsmo}) is taken
equal to 2. The entropy in the one-dimensional case vanishes, as expected, for
$\rho=1$, and the maximum is located at a value of the density which is equal
to 1/2 for monomers ($\alpha=0$) increasing monotonically with $\alpha$ and
approaching 1 in the polymer limit $\alpha \to 1$ (figure \ref{maximum}), 
where the entropy vanishes for all values of $\rho$. The value of the maximum 
entropy is a decreasing function of $\alpha$.

For a given value of $M$, the density which maximizes the entropy is obtained
through he equation
\begin{eqnarray}
(1-\alpha\rho)^{-\alpha}(1-\rho)[\rho(1-\alpha)]^{-(1-\alpha)}=1
\end{eqnarray}
On the square lattice, our calculations indicate that the absolute maximum of
the entropy $s_M(\rho)$ occurs for tetramers ($M=4$). In the one-dimensional
case, the maximum entropy of $s_M(\rho)$ is a monotonically decreasing
function of $M$, the absolute maximum $s_1(1/2)=\ln(2)$ being obtained for
monomers and approaching 0 as $M \to \infty$.

Finally, the problem of $M$-mers confined inside strips of finite width
with closed boundary conditions, is an interesting extension of
earlier work done in the polymer limit \cite{s99}. 

\begin{acknowledgments}
This research was partially supported by the Brazilian agencies CAPES, FAPERJ
and CNPq, whose assistance is gratefully acknowledged. 
\end{acknowledgments}

\end{document}